\documentstyle[prb,aps,twocolumn]{revtex}

\begin{document}
\draft


\newcommand{\be}{\begin{equation}}
\newcommand{\ee}{\end{equation}}
\newcommand{\bea}{\begin{eqnarray}}
\newcommand{\eea}{\end{eqnarray}}
\newcommand{\bi}{\bibitem}

\newcommand{\vp}{\varphi}
\newcommand{\ep}{\epsilon}
\newcommand{\g}{\gamma}
\newcommand{\s}{\sigma}
\newcommand{\D}{\Delta}
\newcommand{\rs}{\rho_s}
\newcommand{\hrs}{\hat{\rho}_s}

\newcommand{\r}{({\bf r})}
\newcommand{\rp}{({\bf r}')}
\newcommand{\rrp}{({\bf r},{\bf r}')}
\newcommand{\rpr}{({\bf r}',{\bf r})}
\newcommand{\xxp}{({\bf x},{\bf x'})}
\newcommand{\xpx}{({\bf x'},{\bf x})}
\newcommand{\xxpr}{({\bf x},{\bf x'};{\bf r})}

\newcommand{\ua}{\uparrow}
\newcommand{\da}{\downarrow}
\newcommand{\la}{\langle}
\newcommand{\ra}{\rangle}
\newcommand{\dg}{\dagger}

\newcommand{\lmt}{\left(\begin{array}{cc}}
\newcommand{\lmf}{\left(\begin{array}{cccc}}
\newcommand{\rma}{\end{array}\right)}
\newcommand{\lvec}{\left(\begin{array}{c}}
\newcommand{\rvec}{\end{array}\right)}

\wideabs{
\title{Spin-distribution functionals and correlation energy of the Heisenberg
model}
\author{Valter L. L\'{\i}bero}
\address{Departamento de F\'{\i}sica e Inform\'atica,
Instituto de F\'{\i}sica de S\~ao Carlos,
Universidade de S\~ao Paulo,\\
Caixa Postal 369, 13560-970 S\~ao Carlos, SP, Brazil}
\author{K. Capelle}
\address{Departamento de Qu\'\i mica e F\'{\i}sica Molecular,
Instituto de Qu\'\i mica de S\~ao Carlos,
Universidade de S\~ao Paulo,\\
Caixa Postal 780, 13560-970 S\~ao Carlos, SP, Brazil}
\date{\today}
\maketitle
\begin{abstract}
We analyse the ground-state energy and correlation energy of the Heisenberg 
model as a function of spin, both in the ferromagnetic and in the 
antiferromagnetic case, and in one, two and three dimensions. First,
we present a comparative analysis of known expressions for the ground-state 
energy $E_0(S)$ of {\it homogeneous} Heisenberg models.
In the one-dimensional antiferromagnetic case we propose an improved
expression for $E_0(S)$, which takes into account Bethe-Ansatz data for
$S=1/2$. Next, we consider {\it inhomogeneous} Heisenberg models (e.g., 
exposed to spatially varying external fields). We prove a Hohenberg-Kohn-like
theorem  stating that in this case the ground-state energy is a functional of 
the spin distribution, and that this distribution encapsulates the entire
physics of the system, regardless of the external fields.
Building on this theorem, we then propose a local-density-type approximation
that allows to utilize the results obtained for homogeneous systems also in
inhomogeneous situations. We conjecture a scaling law for the dependence of the
correlation functional on dimensionality, which is well satisfied by existing
numerical data. Finally, we investigate the importance of the spin-correlation
energy by comparing results obtained with the proposed correlation functional 
to ones from an uncorrelated mean-field calculation, taking as our example a 
linear spin-density wave state.
\end{abstract}

\pacs{PACS numbers: 71.15.Mb, 75.10.Jm, 75.50.Ee, 75.30.Fv}


}

\section{Introduction}
\label{intro}

In this paper we study the ground-state energy, correlation energy and 
related quantities of the Heisenberg model. The {\it homogeneous
Heisenberg model} is defined by the Hamiltonian
\be
\hat{H}_0 = J \sum_{\la ij\ra} \hat{\bf S}_i \cdot \hat{\bf S}_j,
\label{homheis}
\ee
where the $\hat{\bf S}_i$ are spin vector operators satisfying
$\hat{\bf S}_i^2|Sm\ra = S(S+1)|Sm\ra$ and $\hat{S}_{i,z}|Sm\ra = m|Sm\ra$,
and $S$ and $m$ are the spin quantum numbers of the particles under study.
Although in accordance with common terminology the $\hat{\bf S}_i$ are called
spin operators, they really represent total angular momentum and are not 
restricted to be of purely spin origin.
$\la ij\ra$ indicates a sum over nearest neighbors on a lattice of
dimensionality $d$, and $J$ is the spin-spin interaction constant, 
parametrizing the exchange interaction of the underlying microscopic 
Hamiltonian.\cite{mattis,herring,yosida} For antiferromagnetism $J>0$, while
for ferromagnetism $J<0$. This model was originally proposed in 1926 
to explain ferromagnetism in transition metals,\cite{heisenberg,dirac}
but has since then found a large number of other applications to the
magnetic properties of matter.\cite{mattis,herring,yosida} Recent examples are
antiferromagnetic chains in complex oxides and other low-dimensional
magnets \cite{chains} or studies of magnetic effects on crystal-field 
splittings in rare-earth compounds.\cite{valter}

The {\it inhomogeneous Heisenberg model}, characterized by broken translational
symmetry, is obtained by adding a spatially varying magnetic field to 
$\hat{H}_0$,
\be
\hat{H} = J \sum_{\la ij\ra} \hat{\bf S}_i \cdot \hat{\bf S}_j 
 + \sum_i {\bf B}_i \cdot \hat{\bf S}_i,
\label{inhomheis}
\ee 
where ${\bf B}_i$ can either be an externally applied field or an internal
field due to magnetism in the system. A variety of different sources and
manifestations of the inhomogeneity ${\bf B}_i$ has been studied in the
recent literature, often in conjunction with the synthesis and investigation
of real materials, whose magnetic properties are necessarily spatially 
inhomogeneous, but still to
some extent describable by modified Heisenberg models.\cite{chains,inhomrefs}
The interpretation of experimental data for realistic materials in terms of 
the Heisenberg model, in particular in the presence of staggered or otherwise 
spatially varying magnetic fields, must be based on a solid understanding of 
the behaviour of the model (\ref{inhomheis}) in the presence of inhomogeneity. 

The homogeneous Heisenberg model in $d=1$ dimension and for $S=1/2$ has
an exact analytical solution in terms of the Bethe Ansatz,\cite{bethe,hulthen}
but the same Ansatz does not work in higher dimensions, in which no exact
solution is known. It is also hard to generalize to inhomogeneous situations.
In the present paper we combine exact and approximate results obtained within
a variety of different approaches and techniques, to provide a systematic
analysis of the ground-state and correlation energy of the Heisenberg model
both in the ferro and in the antiferromagnetic case and for $d=1,2,3$ 
dimensions.

In Sec.~\ref{gsenergy} we provide a comparative analysis of available
expressions for the ground-state energy of the homogeneous Heisenberg
model as a function of spin $S$, dimenionality $d$, and coupling constant $J$.
This section also contains a proposal for an improved expression that goes 
beyond those available in the literature in taking into account a Bethe Ansatz
result for $S=1/2$. In Sec.~\ref{inhom} we use concepts of density-functional
theory to extend the utility of the homogeneous results discussed in the
preceeding section to inhomogeneous situations. This section includes a proof
of a Hohenberg-Kohn-type theorem for a large class of generalized Heisenberg
models, and a proposal for a simple local-density approximation.
In Sec.~\ref{correlations} we then study the correlation energy in homogeneous
and inhomogeneous Heisenberg models, in order to assess the importance of 
correlations and quantum fluctuations as a function of dimensionality and 
spin. As an explicit example for a physically interesting type of
inhomogeneity we consider a linear spin-density wave, and explore the
differences between a mean-field and a density-functional treatment of
the resulting inhomogeneous spin distribution.

\section{Homogeneous Heisenberg models: Ground-state energy as function of spin}
\label{gsenergy}

This section provides a brief review of what is known about the 
ground-state energy of homogeneous Heisenberg models. Although the 
expressions collected below for $J>0$ and $J<0$ and $d=1,2,3$ are given
at various places in the literature, we have not found a systematic
collection and comparison at one single place. For the convenience of
the reader and future reference we therefore provide such a comparison 
below. We also add a new expression to the list [Eq.~(\ref{valterfit})],
which is a slight improvement on one of the earlier results. 

\subsection{Ferromagnetic case}
\label{fm}

Let us first consider the ferromagnetic case, which is much simpler than
the antiferromagnetic one. At zero temperature all spins are parallel,
and the corresponding spin operators commute with each other. The exact
ground-state energy is then the same as that obtained in the mean-field 
approximation, and is given by 
\be
E_0^{FM}(S,J,z) = J\frac{z}{2}N S^2,
\ee
where $z$ is the number of nearest neighbours, which on linear, square 
and cubic lattices is related to the dimensionality by $d=z/2$ and, as above,
$J<0$ for ferromagnetism. In Fig.~\ref{fig1} we show the resulting curves for 
the energy per site and interaction strength
\be
e_0^{FM}(S,d)={1\over N|J|}E_0^{FM}(S,J,z=2d) = -d S^2,
\ee
for one, two and three dimensions.

\subsection{Antiferromagnetic case}
\label{afm}

The antiferromagnetic (AFM) case is much more complicated than the 
ferromagnetic 
one, because in spite of its name the ground-state of the antiferromagnetic
Heisenberg model is not simply the `antiferromagnetic' state consisting of
alternating spin up and spin down states with respect to a fixed direction
(i.e., the Neel state), but a quantum superposition of states involving also
spins along the perpendicular axes. In the one-dimensional $S=1/2$ case the
structure of the corresponding ground-state for $N\to \infty$ is known exactly 
by means of the Bethe Ansatz.\cite{yosida,bethe,hulthen} The corresponding
ground-state energy is
\bea
E_0^{AFM}(S={1\over 2},J,d=1)
\nonumber \\
= E_0^{FM}(S={1\over 2},J,z=2)-JN \ln 2.
\eea
For the energy per site and interaction strength one obtains from 
this\cite{hulthen}
\bea
e_0^{AFM}(S={1\over 2},d=1)
=\frac{1}{NJ} E_0^{AFM}(S={1\over 2},J,d=1)
\\
=\frac{1}{4}-\ln 2 = -0.44314718.
\eea

Similar exact results are not available in the general case, for arbitrary 
$S$ and $d$. However, a set of useful approximate expressions was derived
in two early papers by Anderson.\cite{anderson1,anderson2}
In the first of these it is shown
by means of a variational argument that the energy of the
antiferromagnetic ground state must lie in the interval \cite{anderson1}
\be
-\frac{z}{2} S^2 \left( 1+\frac{1}{zS}\right) 
\leq e_0^{AFM}(S,z)
\leq -\frac{z}{2} S^2,  
\label{interval}
\ee
where $z=2d$ is again the coordination number of the linear, square or cubic
lattice under study.
A simple estimate is obtained by using the center of this 
interval,\cite{anderson2} i.e.,
\be
e_0^{AFM}(S,z) \approx -\frac{z}{2} S^2 \left( 1+\frac{1}{2 zS}\right),
\label{estimate}
\ee
but the quality of this estimate, which is quite 
good for $d=3$, deteriorates for $d=2$ and $d=1$. A numerical calculation 
based on spin-wave theory leads to the more precise results\cite{anderson2}
\bea
e_0^{AFM}(S,d=1) &=& -S^2 + \left({2\over \pi}-1 \right)S 
\label{pwad1a}
\\
&=& -S^2\left(1+{0.36338\over S} \right) 
\label{pwad1}\\
e_0^{AFM}(S,d=2) &=& -2 S^2 \left(1+{0.158\over S} \right)
\label{pwad2}\\
e_0^{AFM}(S,d=3) &=& -3 S^2 \left(1+{0.097\over S} \right).
\label{pwad3}
\eea
We can assess the quality of these expressions by substituting $S=1/2$ 
in the first of them. The result $e_0^{AFM}(S={1\over 2},d=1)= - 0.43169$
is within $2.6\%$ of the exact Bethe Ansatz value reported above.

Indeed, significant improvement over Eq.~(\ref{pwad1a}) 
has only been obtained recently, with the use of modern computing facilities
and advanced numerical techniques. In such work Lou et al. \cite{lou} used 
(50 years after Ref.~\onlinecite{anderson2}) the density-matrix renormalization
group (DMRG) to calculate corrections to Eq.~(\ref{pwad1a}) for values of $S$
ranging from $1/2$ to $5$ in steps of $1/2$. These authors propose a fit to 
their numerical data which in our present notation reads
\bea
e_0^{AFM,fit1}(S,d=1) = -S^2+\left({2\over \pi}-1\right)S - 0.03262
\nonumber \\
- 0.0030{1\over S} 
- \left(0.338-{0.28 \over S}\right) e^{-\pi S} \cos(2\pi S).
\label{loufit}
\eea
We note that for $S=1/2$, where it predicts $e_0^{AFM,fit1}(S={1\over 2},d=1)= 
-0.516459$, this fit is actually worse than the earlier expression 
(\ref{pwad1}), deviating by $17\%$ from the exact Bethe Ansatz value. 
On the other hand, as shown in Ref.~\onlinecite{lou} by
comparison with DMRG data and other highly precise numerical results, 
the fit is excellent for higher values of $S$. 

In order to obtain a closed expression that can also be applied at $S=1/2$, 
a slight modification to the fit by Lou et al. is sufficient. To this
end we propose the alternative expression
\bea
e_0^{AFM,fit2}(S,d=1) = -S^2+\left({2\over \pi}-1\right)S 
\nonumber \\
- 0.03262 - 0.0030{1\over S} + 0.0015{1\over S^3} 
\nonumber \\
- \left(0.338-{0.28 \over S} + {0.035 \over S^3}\right) e^{-\pi S} \cos(2\pi S),
\label{valterfit}
\eea
which differs from (\ref{loufit}) in the inclusion of two cubic terms in $1/S$.
The value at $S=1/2$ predicted by this expression, 
$e_0^{AFM,fit2}(S={1\over 2},d=1) = -0.446253$, deviates by only $0.7\%$ from 
the exact Bethe Ansatz value.

Figures~\ref{fig2} and \ref{fig3} display the various AFM energy expressions 
collected 
above. In Fig.~\ref{fig2} we compare the rigorous but rather wide interval 
provided by expression (\ref{interval}), with the spin-wave results 
(\ref{pwad1}) - (\ref{pwad3}) and (in $d=1$) the DMRG data of 
Ref.~\onlinecite{lou}. Obviously, on this scale the spin-wave expression
(\ref{pwad1}) already provides an excellent approximation to the DMRG data.
Neither of the two fits (\ref{loufit}) and (\ref{valterfit}) can be
distinguished from the much simpler expression (\ref{pwad1}) on the scale
of the figure. Unfortunately,
no highly precise numerical reference data, similar to the DMRG results 
of Ref.~\onlinecite{lou}, seem to be available in $d=2$ and $d=3$, but
the approximations leading to the simple analytical formulae
given above are expected to work better as $d$ increases. This expectation
is corroborated by noting that the
rigorous interval (\ref{interval}) shrinks with increasing $d$. 

Interestingly, both the numerically highly precise DMRG data of 
Ref.~\onlinecite{lou} (in $d=1$) and the spin-wave expressions
(\ref{pwad1}) to (\ref{pwad3}) (in $d=1,2,3$) systematically lie closer to
the more negative boundary of the interval than to the less negative one, 
and in $d=1$ the DMRG values are still a little closer to this boundary than 
the curve predicted by Eq.~(\ref{pwad1}). This shows that the lower
bound in Eq.~(\ref{interval}) is tighter than the upper one.
The simple estimate (\ref{estimate}), on the other hand, 
by construction falls in the middle of the interval and becomes less
reliable for lower $d$. In the interest of readability we have not displayed
the curves corresponding to this estimate in the figures.

Figure~\ref{fig3} shows that on a smaller scale the differences between the
more precise expressions become important. Here we compare the two fits,
(\ref{loufit}) and (\ref{valterfit}), to the DMRG data. To make the details of
the fits, and the interesting oscillatory structure they display, clearly
visible, we have subtracted the spin-wave expression (\ref{pwad1}), which
is common to both fits. The fit proposed
above, Eq.~(\ref{valterfit}), is slightly inferior to the one developed by
Lou et al., Eq.~(\ref{loufit}), around $S=3/2$, but unlike the latter
recovers the exactly known $S=1/2$ data point to within less 
than one percent.

In the present paper we are mainly concerned with the homogeneous or
inhomogeneous Heisenberg model on linear, square and cubic lattices.
Expressions (or numerical values) for the ground-state energy can also be
derived for many other
variations of the Heisenberg model, such as lattices with helical boundary
conditions,\cite{helical} or with anisotropic interactions.\cite{anisotropic}
Although we do not consider such models in the present paper, many of our
results can be extended to them in a straightforward way.

\section{Inhomogeneous Heisenberg models: Spin-distribution functionals}
\label{inhom}

Based on the analysis of the preceeding section we recommend the use of
expressions (\ref{pwad1}) to (\ref{pwad3}) in calculations requiring
simple expressions for the ground-state energy of the homogeneous 
antiferromagnetic Heisenberg model in one, two, and three dimensions.
In one dimension, where the simple expressions fare worst, either of the
two fits (\ref{loufit}) and (\ref{valterfit}) provides a significant
improvement in accuracy, but only the latter recovers the Bethe Ansatz
value at $S=1/2$. 
However, the utility of any of these expressions is rather limited due to the 
restriction to spatial homogeneity. Externally applied magnetic fields that 
vary in space, calculations of magnetic effects on crystal-field splitting, 
description of nontrivial internal order, etc., require 
use of the {\it inhomogeneous} Heisenberg model.\cite{inhomrefs} 
Unfortunately, if translational invariance is broken the Bethe Ansatz, 
spin-wave theory, DMRG and most other approaches encounter very significant 
computational difficulties.

In the case of {\it ab initio} calculations a many-body technique that has
had considerable success in the application to inhomogeneous systems is
density-functional theory (DFT),\cite{dftbook,rmp,science} but it is not very 
common to apply DFT also to model Hamiltonians. However, following pioneering 
work by Gunnarsson and Sch\"onhammer,\cite{gs} DFT was recently formulated
and applied for the one-dimensional Hubbard model.\cite{lutt,mott,oxford} 
In this section we build on this experience to explore how DFT can become 
a useful tool also in studies of the inhomogeneous Heisenberg model.
To this end we prove, in subsection \ref{hktheo}, a Hohenberg-Kohn-type 
theorem for a wide class of generalized Heisenberg models (of which
the models discussed above are special cases). In subsection \ref{lda}
we then use this theorem and the explicit expressions discussed in 
Sec.~\ref{gsenergy} to construct a simple local-density approximation 
for inhomogeneous Heisenberg models.

\subsection{Hohenberg-Kohn theorem for generalized Heisenberg models}
\label{hktheo}

A first question that must be answered before DFT can be usefully employed is 
what the fundamental variable is. In {\it ab initio} DFT one mostly chooses the
particle density $n\r$ or its spin-resolved counterpart $n_\s\r$,\cite{dftbook,rmp,science}
although other choices are occasionally useful.\cite{vr,ogk,sdwepl} 
In the case of the Hubbard model the basic variable is the site occupation 
number $n_i$.\cite{gs,lutt,mott,oxford} In the present case we propose to use 
the spin vector ${\bf S}_i$, which is the only fundamental dynamical variable 
appearing in the definition of the Heisenberg model.

In the interest of generality, in the present section we consider a
{\it generalized Heisenberg model} of the form
\be
\hat{H}_g = 
\sum_{ij} J_{ij} \hat{\bf S}_i\cdot \hat{\bf S}_j 
+ \sum_i \hat{\bf S}_i \cdot {\bf B}_i.
\label{genheis}
\ee
Unlike in Eqs.~(\ref{homheis}) and (\ref{inhomheis}) the sum in the first
term on the right-hand side is not restricted to nearest neighbours, 
and the interaction $J_{ij}$ can depend in any way on the indices of
the involved sites. In particular, it can extend to next-nearest neighbours
and beyond, or alternate between ferromagnetic and antiferromagnetic 
along some direction in the crystal. Both of these features are
found in realistic magnetic crystals.
This relaxation of constraints on $J$ may appear a considerable complication,
but it turns out that the proof of the Hohenberg-Kohn theorem is essentially 
unaffected by the extra generality. (As pointed out above, we do not consider 
anisotropic Heisenberg models, in which $J$ couples differently to different 
components of ${\bf S}$, in this paper, but the generalization of the 
theorem to this case is straightforward.) 

Following the steps of Hohenberg and Kohn, we now consider two Hamiltonians 
with same interaction $J_{ij}$, but exposed to two different magnetic 
fields ${\bf B}_i$ and ${\bf B}_i'$. Thus
\bea
\hat{H} \Psi=
\left[\sum_{i,j} J_{ij} \hat{\bf S}_i\cdot \hat{\bf S}_j +
\sum_i \hat{\bf S}_i \cdot {\bf B}_i \right] \Psi = E_0 \Psi
\\
\hat{H}' \Psi'=
\left[\sum_{i,j} J_{ij} \hat{\bf S}_i\cdot \hat{\bf S}_j +
\sum_i \hat{\bf S}_i \cdot {\bf B}'_i \right] \Psi' = E_0' \Psi',
\eea 
where $E_0$ and $E_0'$ are the ground-state energies in the 
fields ${\bf B}_i$ and ${\bf B}_i'$, and $\Psi$ and $\Psi'$ are the
corresponding ground-state wave functions.

As a consequence of the variational principle we have the inequality
\be
E_0=\la \Psi | \hat{H} | \Psi \ra < \la \Psi' | \hat{H} | \Psi' \ra,
\ee
since $\Psi'$ is not the ground-state wave function belonging to
$\hat{H}$ (assumed nondegenerate). By adding and subtracting the term 
$\sum_i {\bf \hat{S}}_i \cdot {\bf B}'_i$ on the right-hand side, the 
inequality becomes
\be
E_0=\la \Psi | \hat{H} | \Psi \ra 
< \la \Psi' |\hat{H}' | \Psi' \ra
+\sum_i\la \Psi' |\hat{\bf S}_i \cdot ({\bf B}_i-{\bf B}'_i)
| \Psi' \ra.
\ee
Here the first term on the right-hand side is just the ground-state energy 
$E_0'$, of Hamiltonian $\hat{H}'$. With the abbreviations
$\Delta {\bf B}_i = {\bf B}_i-{\bf B}'_i$,
${\bf S}_i = \la \Psi | \hat{\bf S}_i | \Psi \ra$ and
${\bf S}'_i = \la \Psi' | \hat{\bf S}_i | \Psi' \ra$
the preceeding equation then becomes
\be
E_0 < E_0' + \sum_i {\bf S}'_i \cdot \Delta {\bf B}_i.
\label{contra1}
\ee

Now we repeat the same argument starting with the Hamiltonian $\hat{H}'$.
The variational principle guarantees that
\be
E_0'=\la \Psi' | \hat{H}' | \Psi' \ra < \la \Psi | \hat{H}' | \Psi \ra.
\ee
By adding and subtracting the term $\sum_i {\bf \hat{S}}_i \cdot {\bf B}_i$,
we obtain, in the same way as before,
\be
E_0' < E_0 - \sum_i {\bf S}_i \cdot \Delta {\bf B}_i.
\label{contra2}
\ee
Addition of Eqs.~(\ref{contra1}) and (\ref{contra2}) leads to
\be
E_0+E_0' < E_0+E_0' + \sum_i ({\bf S}'_i-{\bf S}_i) \Delta {\bf B}_i.
\ee
If we now assume that 
${\bf S}'_i={\bf S}_i$, i.e., that the two spin distributions 
corresponding to the two different wave functions $\Psi$ and $\Psi'$ are 
identical, then the previous equation reduces to the contradiction
\be
E_0+E_0' < E_0+E_0'.
\ee

This contradiction shows that two distinct nondegenerate ground states can 
never lead to the same spin distribution. Hence, given some arbitrary spin 
distribution
${\bf S}_i$ there is {\it at most one} nondegenerate wave function which 
gives rise to it. In other words: the spin distribution uniquely
determines the wave function. This means that the wave function is a
functional\cite{footnote1} of the spin distribution, i.e.,
$\Psi=\Psi[{\bf S}_i]$.
This is the statement of the Hohenberg-Kohn theorem for the Heisenberg
model. 

For completeness we mention that the above proof by contradiction, patterned
after the one first presented by Hohenberg and Kohn in the {\it ab initio}
case,\cite{hk} is not the only
possible one. The constrained-search technique of Levy \cite{levy} and Lieb
\cite{lieb} is also easily adapted to the present case. The ground-state
wave function in this approach is uniquely defined by its spin distribution 
as the wave function that minimizes $\la \Psi | \hat{H} | \Psi \ra$
and reproduces ${\bf S}_i$. This minimization defines the functional
\be
E_{LL}[{\bf S}_i] = \min_{\Psi\to{\bf S}_i} \la \Psi | \hat{H} | \Psi \ra.
\label{llfunc}
\ee
whose minimum is the ground-state energy.

An immediate consequence of either formulation of the proof is that the 
ground-state expectation value of any observable $\hat{O}$, is also a 
functional of the spin distribution, defined via
\be
O[{\bf S}_i]= \la \Psi[{\bf S}_i] | \hat{O} | \Psi[{\bf S}_i] \ra,
\label{ofunctional}
\ee
and this functional is the same regardless of the strength and direction
of the magnetic field ${\bf B}_i$, i.e., it is universal with respect to 
external fields. Note that the theorem applies to any ground-state observable. 
For example, it implies that also all multi-spin correlation functions of 
the general form
\be 
C_{n,n+1,n+2,\ldots}:=
\la \Psi| \hat{\bf S}_n \hat{\bf S}_{n+1} \hat{\bf S}_{n+2} \ldots | \Psi \ra
\ee
are uniquely determined by the single-spin expectation value 
${\bf S}_n = \la \Psi | \hat{\bf S}_n | \Psi \ra$. This is trivially true
for the nearest-neighbor correlation function $C_{n,n+1}^{hom}$ of a
homogeneous one-dimensional system, which as a consequence of the definition 
of the homogeneous Hamiltonian (\ref{homheis}) is simply given by
\be
C_{n,n+1}^{hom} = \frac{1}{NJ} E_0(S),
\ee
but the above proof guarantees that the more complicated correlation
functions involving more than two spins and/or spatially inhomogeneous 
spin distributions are, in principle, also functions of ${\bf S}_i$ only.

\subsection{Local-density approximation}
\label{lda}

Another consequence of this Heisenberg-model formulation of the Hohenberg-Kohn
theorem is that the model's ground-state energy and spin distribution can be 
obtained by application of the variational principle to spin distributions, 
instead of wave functions. In complex situations this can be a 
major simplification, 
but to extract this information is, of course, still highly nontrivial. 
The most straightforward thing to do would be to set up an approximation for
the total energy of the system under study as a functional of the
spin distribution, and to minimize with respect to ${\bf S}_i$.
In {\it ab initio} DFT this is not the preferred way to proceed because it
turns out to be hard to conceive good density functionals for the kinetic 
energy. In practical applications of {\it ab initio DFT} one therefore
commonly employs an indirect minimization scheme, leading to the 
widely used Kohn-Sham equations. Although this could also be done in
the present case, there is no need for introducing a Kohn-Sham system
for the Heisenberg model, since there is no kinetic-energy term in the 
first place. Direct minimization of total-energy expressions seems much
more convenient (and more analogous to the way the model is usually
treated in statistical physics) than indirect minimizations.
In order to explore how such a direct minimization can proceed we first
construct, in this subsection, a local-density approximation for the simpler 
Heisenberg models discussed in the preceeding section. 

Let us write the two contributions to the ground-state energy of the 
inhomogeneous Heisenberg model as $E_J$ and $E_B$, which are the
ground-state expectation values of the first and second term on the right-hand 
side of Eq.~(\ref{inhomheis}), respectively.
The mean-field approximation for $E_J$ yields
\bea
E^{MF}_0[{\bf S}_i] = J \sum_{ij} {\bf S}_i\cdot {\bf S}_j
+ \sum_i {\bf S}_i \cdot {\bf B}_i 
\nonumber \\
\equiv E^{MF}_J[{\bf S}_i] + E_B[{\bf S}_i],
\label{emfdef}
\eea
where, as above, ${\bf S}_i = \la \Psi | \hat{\bf S}_i | \Psi \ra$.
In Sec.~\ref{correlations} we quantify the error made by the neglect of
correlation effects arising from use of $E^{MF}_J[{\bf S}_i]$ in place
of $E_J[{\bf S}_i]$.

A guideline for the construction of better functionals than 
$E^{MF}_J[{\bf S}_i]$ is provided by {\it ab initio} DFT \cite{dftbook} or 
recent work on the Hubbard model.\cite{lutt,mott,oxford}
In the former, the total energy of an arbitrarily inhomogeneous system is 
written as
\be 
E_0[n\r] = T_s[n\r] + E_H[n\r] + E_v[n\r] + E_{xc}[n\r],
\label{dften}
\ee
where $T_s$ is the noninteracting kinetic energy, 
\be
E_H[n\r]=\frac{1}{2}\int d^3r \int d^3r'\, \frac{n\r n\rp}{|{\bf r}-{\bf r}'|}
\ee
the Hartree energy, and $E_v$ the potential energy arising from the external 
field $v\r$. The local-density approximation (LDA) for the exchange-correlation 
($xc$) energy is
\be
E_{xc}[n\r] \approx E_{xc}^{LDA}[n\r] 
= \int d^3r\, e_{xc}(n)|_{n\to n\r}.
\label{abinitiolda}
\ee
This expression locally substitutes the $xc$ energy of the
inhomogeneous system by the one of a homogeneous system of same density. 
The necessary input expression for $e_{xc}(n)$ is obtained by
subtracting the Hartree and noninteracting kinetic energy from the 
ground-state energy of the homogeneous system, $e_0(n)$.

In the present context we write, in analogy to Eq.~(\ref{dften}),
\be
E_0[{\bf S}_i] = E^{MF}_J[{\bf S}_i] + E_B[{\bf S}_i] + E_c[{\bf S}_i],
\label{e0lda}
\ee
where $E^{MF}_J$ is defined in Eq.~(\ref{emfdef}),
$E_B$ is the potential energy arising from
the external field ${\bf B}_i$, and $E_c$ is by definition the
difference between the mean-field result and the correct one, i.e., the
correlation energy. (There is no Heisenberg-model counterpart to the kinetic 
energy term, and we avoid the expression `exchange-correlation
energy' because in common terminology the entire Heisenberg Hamiltonian
is due to `exchange'.)

To obtain an explicit scheme we now propose to approximate, in analogy to 
(\ref{abinitiolda}),
\be
E_c[{\bf S}_i] \approx E_c^{LDA}[{\bf S}_i] 
= \sum_i e_c(S)|_{S\to |{\bf S}_i|},
\label{ldafunc}
\ee
where $e_c(S)$ is obtained by subtracting the mean-field energy, $-d S^2$,
from the homogeneous expressions for $e_0(S)$ discussed in Sec.~\ref{gsenergy}.
As an explicit example, the LDA approximation for the correlation energy of 
an {\it inhomogeneous} antiferromagnetic Heisenberg model in one dimension 
becomes
\be
E^{LDA}_{c,AFM,d=1}[{\bf S}_i] = 
J \left({2\over\pi}-1 \right) \sum_i |{\bf S}_i|,
\label{ldafuncd1}
\ee
where we used Eq.~(\ref{pwad1a}) for $e_0(S)$. Of course Eqs.~(\ref{loufit})
or (\ref{valterfit}) can be used in the same way in $d=1$, and 
Eqs.~(\ref{pwad2}) and (\ref{pwad3}) in $d=2$ and $d=3$, respectively.
The full ground-state energy is then for any $d$ approximated as
\be
E_0[{\bf S}_i] \approx 
E^{LDA}_0[{\bf S}_i] =
E^{MF}_J[{\bf S}_i] + E_B[{\bf S}_i] + E^{LDA}_c[{\bf S}_i].
\ee

Clearly Eq.~(\ref{ldafunc}) is a rather simple approximation, whose quality
may vary widely depending on the circumstances (e.g., values of $J$ and $d$,
or spatial dependence of ${\bf B}_i$). At present it is motivated mainly by 
the considerable practical success of its counterpart in {\it ab inito} 
DFT,\cite{dftbook,rmp,science} and by the encouraging results obtained 
recently with a Bethe-Ansatz based LDA for inhomogeneous Hubbard
models.\cite{lutt,mott,oxford} 
It is clear, however, that the LDA contains essential correlation effects not 
accounted for by the mean-field expression (\ref{emfdef}). In spite of the 
extra term, minimization of (\ref{e0lda}) with (\ref{ldafunc}) is no more 
complicated than that of (\ref{emfdef}). Eq.~(\ref{ldafunc}) thus shows one 
way in which the expressions listed in Sec.~\ref{gsenergy} for homogeneous 
systems can be applied to inhomogeneous situations. 
A simple application of these ideas is worked out in Sec.~\ref{inhomcorr}.

\subsection{A scaling hypothesis}
\label{scaling}

An interesting feature of the functionals obtained by combining 
Eq.~(\ref{ldafunc}) with the explicit formulae of Sec.~\ref{gsenergy}
is that the resulting expressions
depend explicitly on the interaction $J$ and the dimensionality $d$.
The dependence of the {\it ab initio} functionals on these parameters is
not well known, and in particular the $d$-dependence of the functional is
still subject of many ongoing investigations.\cite{lowdimdft} Even in the
much simpler case of the Hubbard model it is only the interaction 
dependence which is featured explicitly in the available functionals,\cite{gs,lutt,mott,oxford} whereas the dependence on dimensionality is essentially 
unknown. In this context it may be useful to have, for the Heisenberg model,
a number of simple expressions that depend explicitly on 
dimensionality and interaction, so that the role of these parameters 
in the functional can be explored in a simplified environment. 

As an explicit example, we consider scaling properties of the
functional as a function of dimensionality $d$.
The Hartree-like term $\propto S^2$ in Eqs.~(\ref{pwad1}) to (\ref{pwad3})
clearly scales linearly with $d$. Hence
\be
e^{MF}_J(d) = d\, e^{MF}_J(d=1).
\ee

Interestingly, the correlation energy contribution $\propto S$ also obeys
a similar, albeit less obvious, scaling law. From the explicit expressions
(\ref{pwad1}) to (\ref{pwad3}) for $e_0^{AFM}(d)$ one obtains for the 
relation between the ratio of the correlation energies and the ratio
of the dimensionalities 
\be
\frac{e_c(d=2)}{e_c(d=1)} = 0.870 = \left({2\over 1}\right)^x
\ee
and
\be
\frac{e_c(d=3)}{e_c(d=2)} = 0.921 = \left({3\over 2}\right)^y,
\ee
where $x$ and $y$ are exponents to be determined. Numerically one finds
$x=-0.201$ and $y=-0.203$. The near-equality of these two exponents among
each other and to the integer fraction $-1/5$ leads us to conjecture the 
following {\it dimensional scaling law}:
\be
e_c(d) = d^{-\eta} e_c(d=1),
\label{scalinglaw}
\ee
where the scaling exponent $\eta = 1/5$. This scaling law accounts for the
numbers in Eqs.~(\ref{pwad1}) to (\ref{pwad3}) to within $\approx 10^{-3}$.

Of course, at present the scaling law (\ref{scalinglaw}) is only a conjecture, 
but one that is consistent with the numbers of spin-wave theory. It also 
correctly predicts that as $d\to\infty$ the correlation energy vanishes,
leaving behind only the mean-field contribution to the total
energy. Since very little is known about the dimension dependence of
density functionals we cannot say at present whether the existence of such 
a law is a mere coincidence, a particular property of the Heisenberg model, 
or a general phenomenon, but we hope that our observation of dimensional 
scaling stimulates further research along these lines.

One practical use that can be made of Eq.~(\ref{scalinglaw}) is to convert
an approximate functional obtained for some value of $d$ into one for another
dimensionality. Counterparts to this property for other Hamiltonians would
be interesting not only for {\it ab initio} calculations (in which many
results are known for $d=3$, but much less in $d=2$ or $d=1$),\cite{lowdimdft}
but also in the case of the Hubbard model, in which the LDA functional is 
known only for $d=1$.\cite{lutt,mott,oxford} 

\section{Correlation energy of the antiferromagnetic Heisenberg model}
\label{correlations}

In this section we apply the results obtained above to a study of the
correlation energy of the antiferromagnetic Heisenberg model. 
In Sec.~\ref{homcorr} we compare the homogeneous expressions of 
Sec.~\ref{gsenergy} with their mean-field approximation, to assess the
importance and behaviour of the correlation energy. 
In Sec.~\ref{inhomcorr} we study a physically interesting inhomogeneity, 
a spin-density wave, with the LDA functional (\ref{ldafuncd1}).

\subsection{Homogeneous system}
\label{homcorr}

As in the previous section we define the correlation energy as
the difference between the total ground-state energy and its mean-field
approximation. For a homogeneous system on a linear, square or cubic lattice
the latter yields
\be
E_0^{MF}(S,J,d) = - J d N S^2
\ee
and thus $e_0(S)=-d S^2$.
In the inset of Fig.~\ref{fig4} we plot the difference between this value and 
the total energy expressions Eqs.~(\ref{pwad1}), (\ref{pwad2}) and 
(\ref{pwad3}). For one dimension we also plot the difference 
between the mean-field result and the more precise expression (\ref{valterfit}).
These differences represent the {\it absolute} size of the correlation energy.
In the main part of Fig.~\ref{fig4} we display the {\it relative} size of 
the correlation energy as compared to the mean-field energy.
Several conclusions can be drawn from inspection of these curves:

(i) The inset of Fig~\ref{fig4} shows that the absolute size of the correlation
energy 
increases towards larger spins. This seems counterintuitive, because larger 
spins should more closely mimick the classical limit, in which there are no 
quantum fluctuations and the mean-field approximation becomes exact in the 
ground state. However, as shown in the main figure, the {\it relative} weight
of the correlation energy as compared to the mean-field energy decreases
towards larger spins. Interestingly, the naive expectation that correlations
should become less important near the classical limit is thus only true
in relative terms, but not in absolute ones.

(ii) On similar grounds one would expect that correlations become less
important for larger dimensionality. This is confirmed both by the main
figure and the inset, showing that correlations decrease in absolute size and 
relative to the mean-field energy as $d$ increases. The way the $d$-dependence
approaches the classical limit is thus qualitatively different from the
way the spin-dependence does.

(iii) The mean-field energy is not reliable for any dimensionality $d\leq 3$ 
and $S<5$, leading to errors that can be larger than $J$ in absolute size 
and larger than $50\%$ in relative terms. This observation puts tight limits 
on the reliability of this rather widely used approximation.

(iv) The improved treatment of correlations in $d=1$, represented by the curves
labeled `DMRG fit', does not invalidate conclusions (i), (ii), and (iii), 
obtained on the basis of the spin-wave expressions (\ref{pwad1}) to 
(\ref{pwad3}). However, both the main figure and the inset of Fig.~(\ref{fig4})
show that it enhances the importance of correlations with respect to the 
mean-field values, as compared to the simpler expressions.

\subsection{Example of an inhomogeneous system: a linear spin-density wave}
\label{inhomcorr}

As an example of
a truly inhomogeneous situation, to which our LDA functional (\ref{ldafunc})
can be applied, we now consider a simple but physically interesting 
inhomogeneity, namely a spin-density wave (SDW) imposed by an external field
on a chain with antiferromagnetic coupling.

We model the SDW state by taking
\be
{\bf S}_n =  S {\bf u}_x \cos \phi_n,
\label{sdw}
\ee 
where $\phi_n= 2\pi(n-1)/\lambda$ and ${\bf u}_x$ denotes the unit vector in
the $x$ direction. This choice describes a linear SDW of amplitude $S$ and wave
length $\lambda$, polarized along the $x$-direction. (The lattice is taken to
be a chain along the $z$-direction.) The corresponding mean-field energy is
\bea
E_0^{MF}[{\bf S}_i] = J \sum_{n=1}^{N-1} {\bf S}_n \cdot {\bf S}_{n+1}
+ \sum_{n=1}^{N} {\bf S}_n \cdot {\bf B}_n 
\nonumber \\
\equiv E_J^{MF}[{\bf S}_i] + E_B[{\bf S}_i],
\label{mfdef}
\eea
where $N$ is the number of lattice sites, and ${\bf B}_n$ is a magnetic field 
that can be thought of as either externally applied [thus forcing the system 
into a state with spin distribution (\ref{sdw})] or generated 
self-consistently, or a combination of both.
The LDA approximation for the ground-state energy is, on the other hand,
\bea
E_0^{LDA}[{\bf S}_i] 
= E_0^{MF}[{\bf S}_i] + E_c^{LDA}[{\bf S}_i]
\nonumber \\
= E_J^{MF}[{\bf S}_i] + E_B[{\bf S}_i] + 
\sum_{n=1}^{N} e_c^{AFM}(S)|_{S\to |{\bf S}_i|}
\nonumber \\
\equiv E_J^{LDA}[{\bf S}_i] + E_B[{\bf S}_i],
\label{ldaapprox}
\eea
where we use, for simplicity, Eq.~(\ref{ldafuncd1}) for $E_c^{LDA}[{\bf S}_i]$.

For the given spin distribution (\ref{sdw}) we now compare the predictions
of the mean-field and LDA expressions for the interaction energy $E_J$.
Note that this is not a self-consistent calculation, but a comparison of
the two expressions (\ref{mfdef}) and (\ref{ldaapprox}) for a fixed 
distribution specified by (\ref{sdw}).
In Fig.~\ref{fig5} we plot $E_J^{MF}$, $E_J^{LDA}$ and $E^{LDA}_c$ as functions 
of $\lambda$, the wave length of the SDW. Since the term $E_B$, arising
from the magnetic field, is the same in both approximations we only display
the interaction energy $E_J$. The presence of an external field with symmetry
different from the one of the ground-state of the unperturbed 
antiferromagnetic system gives rise to a rich physics. Particularly, we note 
the following points:

(i) Addition of the correlation energy to the mean-field result lowers
the total energy considerably. This is, qualitatively, the same behaviour
we found previously in homogeneous systems and illustrates again the
importance of going beyond the mean-field expression for the energy.

(ii) As $\lambda\to \infty$ the SDW approaches a ferromagnetic spin
configuration. Since the unperturbed homogeneous model is antiferromagnetic
this state is energetically unfavored, and corresponds to a maximum of
the $E$ versus $\lambda$ curves. In the opposite limit the spatial
modulation of the SDW can take local advantage of the AFM tendency of the
underlying homogeneous system. This leads to a lowering of the energy as 
$\lambda \to 0$.

(iii) Once $\lambda$ is larger than approximately $10$ lattice constants the
$E$ versus $\lambda$ curves saturate. We interpret this in terms of the 
correlation length of the antiferromagnetic model by noting that once the
SDW modulation takes place on a scale larger than a few correlation lengths,
the system will be relatively insensitive to further approximation to the
ferromagnetic state.

(iv) The overall downshift of the LDA 
curve compared to the mean-field one implies that the ferromagnetic 
configuration is energetically less unfavorable in the former approximation
than in the latter. Physically this is reasonable, because the correlations
accounted for by $E_c$ break up the rigid AFM pattern of the Neel state found
in the mean-field approximation, and replace it by a complex ground-state
involving spins along all three directions in space. 

\section{Summary and outlook}

Density-functional theory is commonly applied to the {\it ab initio}
Hamiltonian, in the context of electronic structure calculations of 
molecules or solids.\cite{dftbook,rmp,science} Applications to model 
Hamiltonians are rare, although they can be useful both for the analysis 
of these models in the presence of inhomogeneity (broken translational
invariance), and for further development of DFT. The present work on the 
Heisenberg model serves to exemplify these two complementary aspects.

The Heisenberg Hamiltonian is much simpler than the Hubbard
Hamiltonian, which describes the charge-degrees of freedom in addition
to the spin ones, or than the {\it ab initio} Hamiltonian involving the
long-range Coulomb interaction between charges in a real crystal.
The Heisenberg model may therefore be considered a simplified environment
in which concepts and methods of DFT can be tested and analysed. An interesting
example is the study of the interaction and dimension dependence of $xc$
functionals, about little is known in the {\it ab initio} case.

On the other hand, the LDA for inhomogeneous Heisenberg models is no more 
complicated, formally, than the mean-field approximation, but it accounts 
by construction for essential correlation effects missed by the latter. 
Analysis of the correlation energy of homogeneous and inhomogeneous 
Heisenberg models, as function of spin and dimensionality, illustrates that 
such effects are crucial for a quantitative description of the ground-state. 
The above LDA functional may thus be useful in studies of the behaviour of 
the Heisenberg model in spatially varying external fields or in the presence
of internal magnetic order that breaks translational symmetry.

The simple LDA-type correlation functional based on Eqs.~(\ref{ldafunc})
and (\ref{pwad1}) is seen to produce significant quantitative change
and qualitative improvement over the mean-field approximation, at very
little extra computational cost. This observation encourages us to envisage
more complex applications of this functional, such as to impurity states 
in the Heisenberg model.
An extension of the present work to a study of the thermodynamics of
inhomogeneous Heisenberg models (employing the $T>0$ formulation of 
DFT\cite{mermin}) is also planned for the future.

{\bf Acknowledgments:} {\it This work was supported by FAPESP.
We thank L.~N.~Oliveira and N.~A.~Lima for useful discussions.}

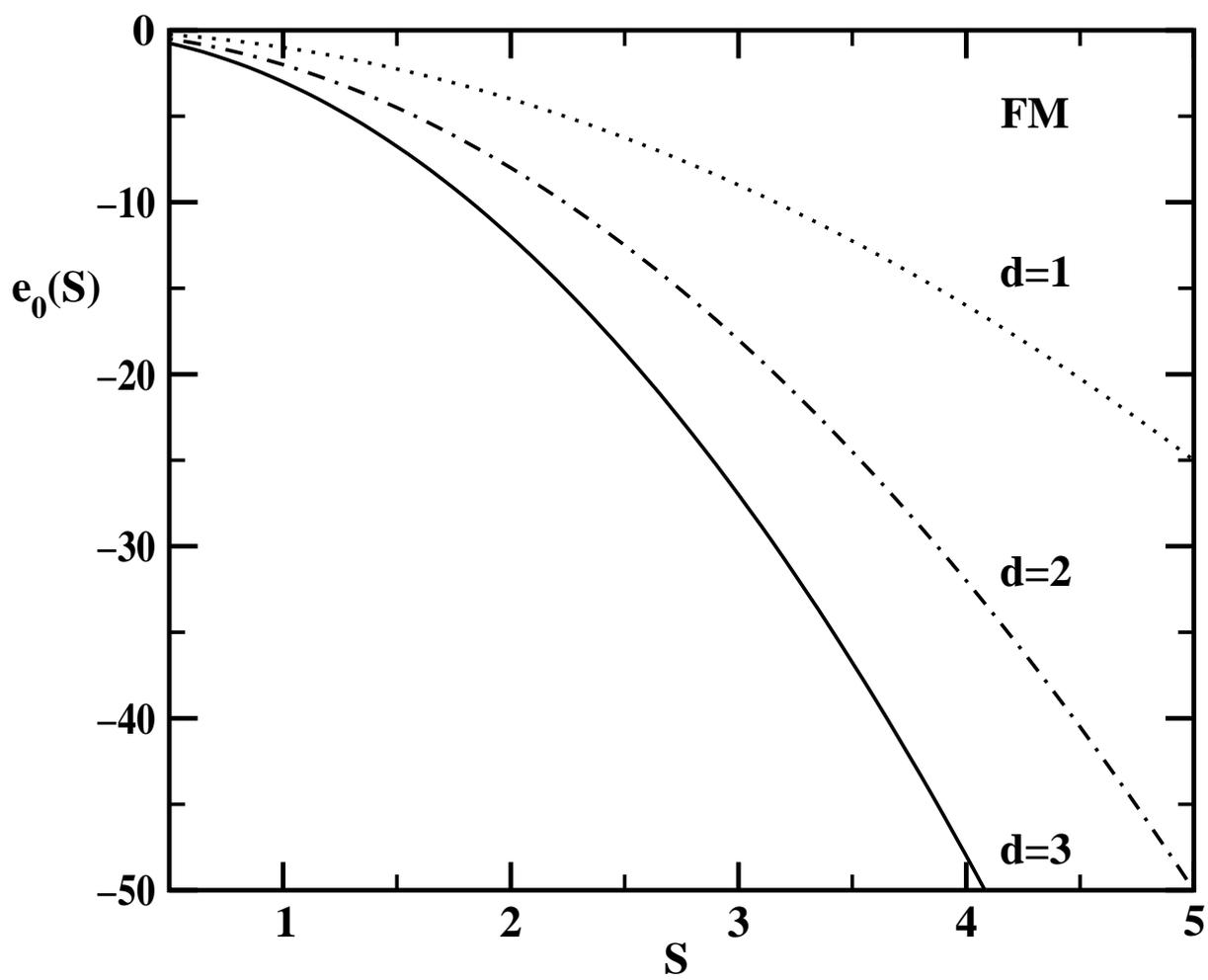
\begin{figure}
\caption{Per-site ground-state energy of the homogeneous
ferromagnetic Heisenberg model in one, two and three dimensions.}
\label{fig1}
\end{figure}

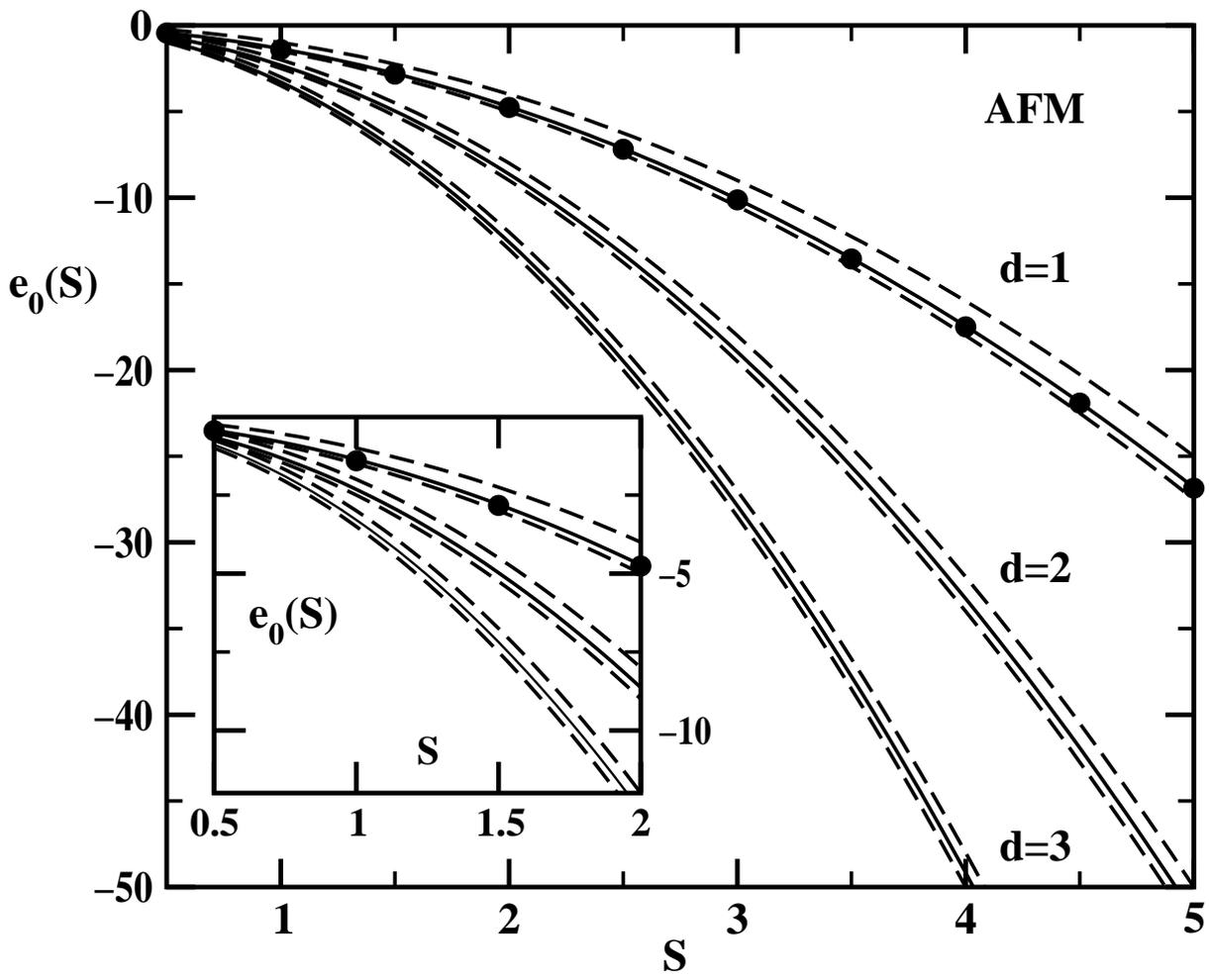
\begin{figure}
\caption{Per-site ground-state energy of the homogeneous antiferromagnetic
Heisenberg model in one, two and three dimensions. 
Dashed lines: interval according to Eq.~(\ref{interval}). 
Full lines: spin-wave theory expressions (\ref{pwad1}), (\ref{pwad2}) and 
(\ref{pwad3}), respectively. 
Full circles (in one dimension only):
DMRG data of Ref.~\protect\onlinecite{lou}. 
The inset is a zoom into the low spin region of the main figure.}
\label{fig2}
\end{figure}

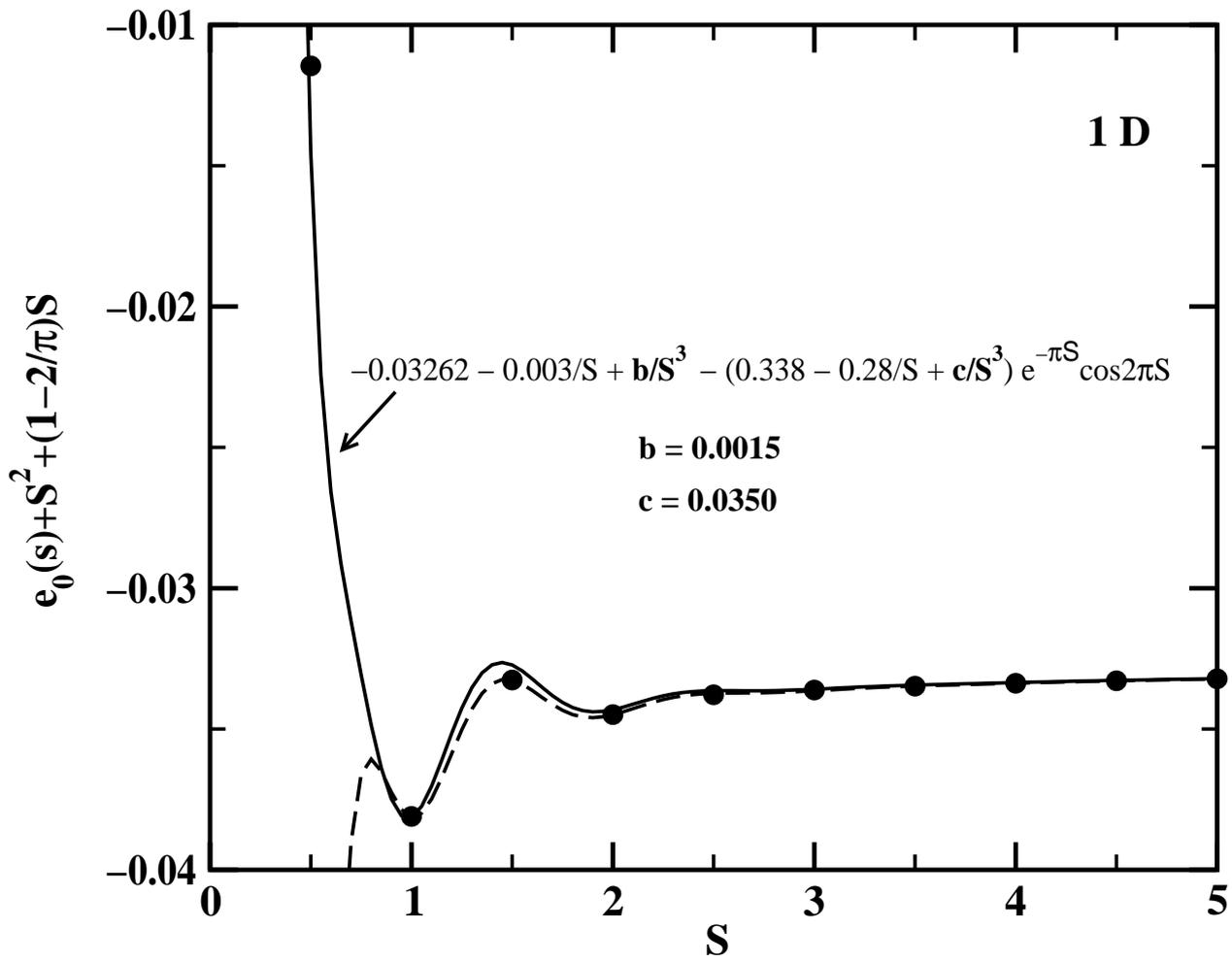
\begin{figure}
\caption{Same as in Fig.~\ref{fig2}, but on a reduced scale.
On the vertical axis the expression (\ref{pwad1}) has been subtracted to
display only the corrections to that result. Full circles: DMRG data of 
Ref.~\protect\onlinecite{lou}, dashed line: fit proposed
in that reference, present Eq.~(\ref{loufit}), full line: fit proposed
here, Eq.~(\ref{valterfit}).}
\label{fig3}
\end{figure}

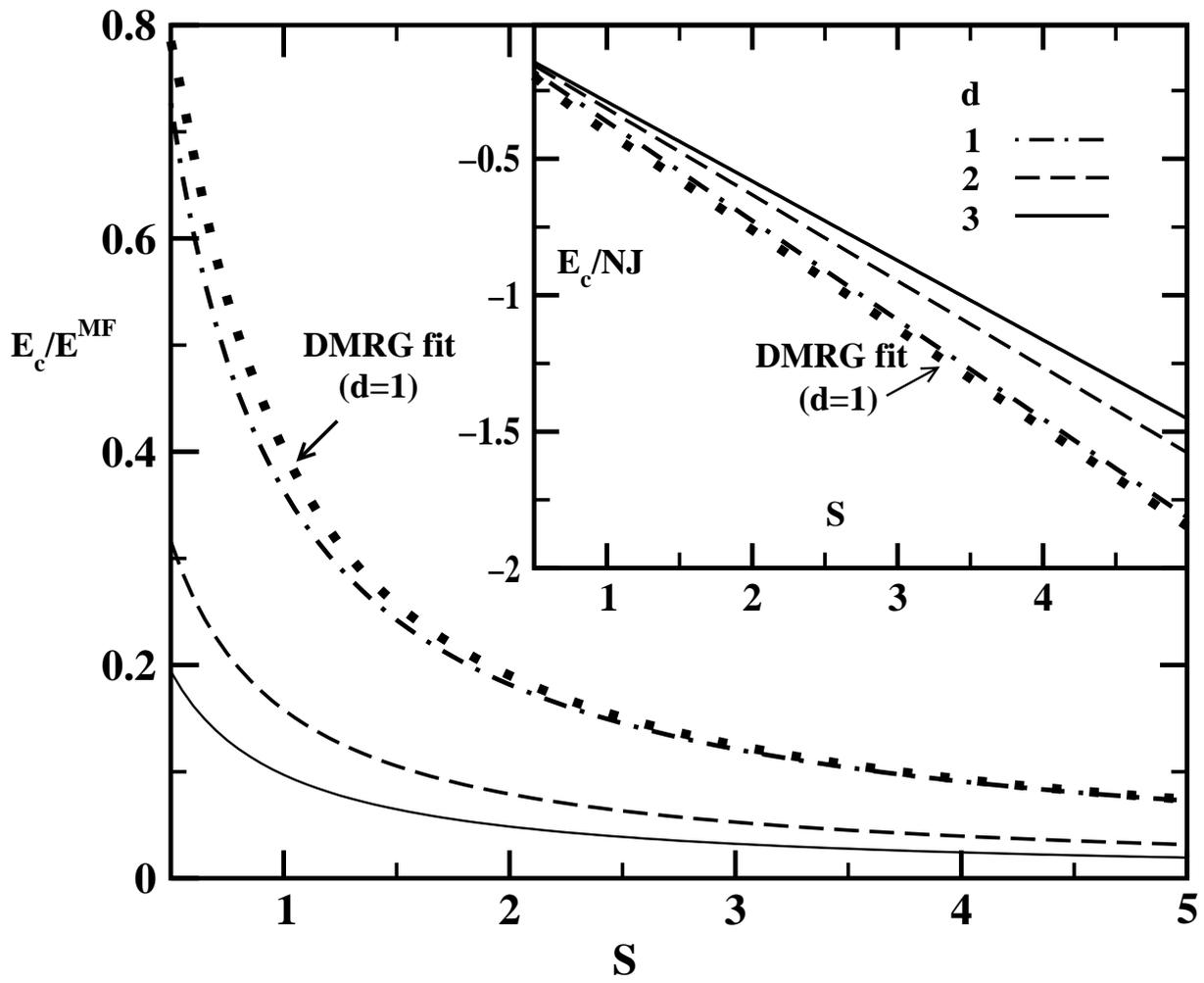
\begin{figure}
\caption{Correlation energy of the antiferromagnetic homogeneous
Heisenberg model in one, two and three dimensions, as a function of spin.
The curves labeled $d=1,2,3$ are obtained by subtracting the mean-field energy
from the expressions (\ref{pwad1}) - (\ref{pwad3}), and the one labeled
`$d=1$ DMRG-fit' is obtained by using our fit Eq.~(\ref{valterfit}) instead
of Eq.~(\ref{pwad1}). Main figure: correlation energy relative to mean-field 
energy. Inset: absolute size of correlation energy.}
\label{fig4}
\end{figure}

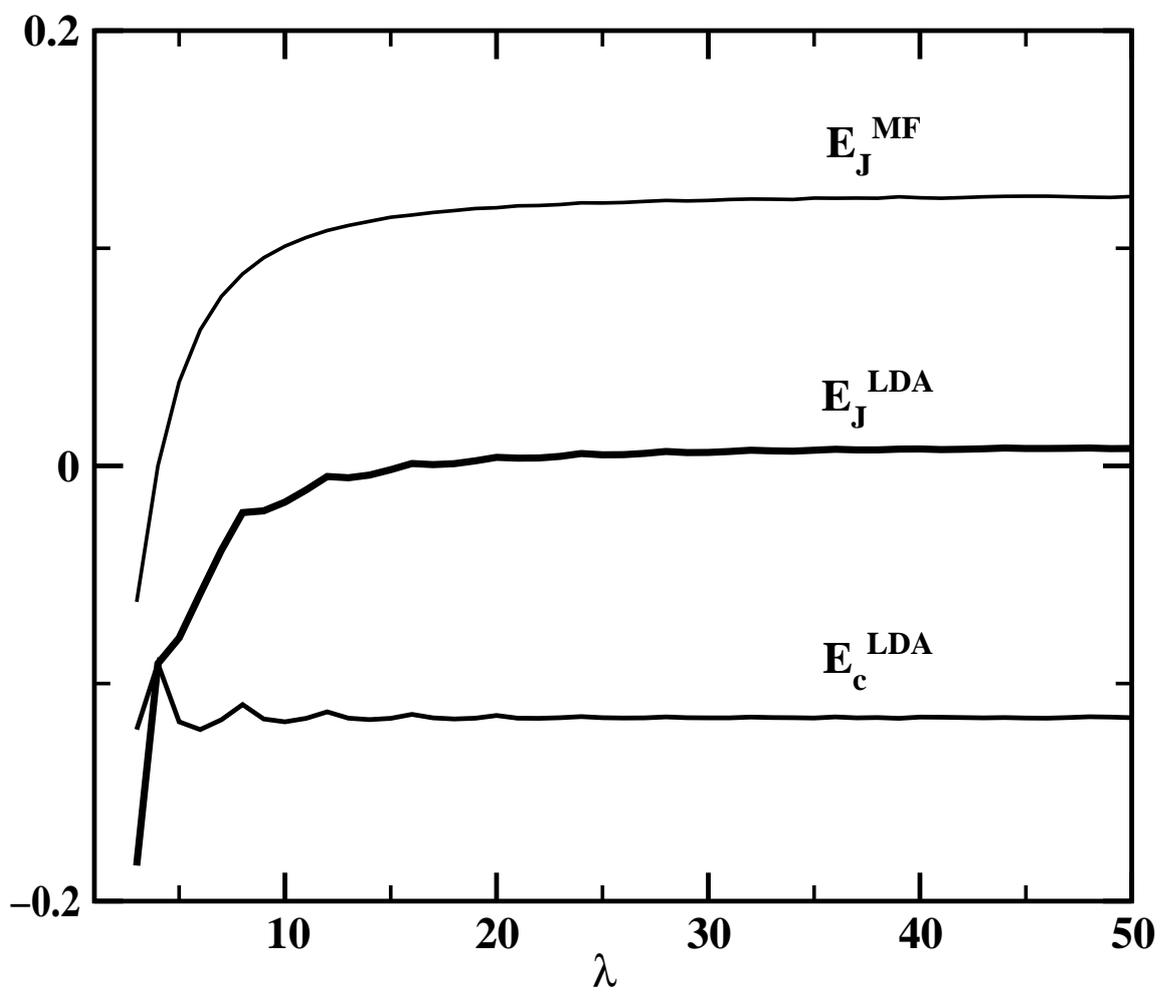
\begin{figure}
\caption{Interaction energy $E_J$ of a spin-density wave imposed on an 
antiferromagnetic chain, as a function of wave length of the spin modulation.
Upper curve: mean-field approximation, defined by Eq.~(\ref{mfdef}). 
Lower curve: local-density approximation for the correlation energy, as given 
by Eq.~(\ref{ldafuncd1}).
Middle curve: LDA result for interaction energy, obtained as sum of the
two other curves.}
\label{fig5}
\end{figure}
\end{document}